\documentclass[aps,pre,superscriptaddress,twocolumn]{revtex4-1}
\usepackage{graphicx}
\usepackage{times}
\usepackage{tabularx}
\usepackage{amsmath}
\usepackage{amsthm}
\usepackage{amssymb}
\usepackage{amsbsy}
\usepackage{verbatim}

\usepackage{color}

\usepackage{float}
\usepackage{hyperref}

\newcommand{\sgn}{\text{sign}}

\newcommand{\be}{\begin{equation}}
\newcommand{\ee}{\end{equation}}
\newcommand{\ba}{\begin{eqnarray}}
\newcommand{\ea}{\end{eqnarray}}
\newcommand{\up}{\uparrow}
\newcommand{\dn}{\downarrow}

\usepackage{hyperref}
\hypersetup{
        colorlinks=true,
}

\begin{document}
\title{Coulomb Blockade in Fractional Topological Superconductors}
\author{Younghyun Kim}
\affiliation{Physics Department, University of California,  Santa Barbara, CA 93106, USA}

\author{David J. Clarke}
\affiliation{Condensed Matter Theory Center, Department of Physics, University of Maryland, College Park}
\affiliation{Joint Quantum Institute, University of Maryland, College Park, Maryland 20742, USA}
\affiliation{Station Q, Maryland}

\author{Roman M. Lutchyn}
\affiliation{Station Q, Microsoft Research, Santa Barbara, California 93106-6105, USA}

\date{\today}

\begin{abstract}
We study charge transport through a floating mesoscopic superconductor coupled to counterpropagating fractional quantum Hall edges at filling fraction $\nu=2/3$. We consider a superconducting island with finite charging energy and investigate its effect on transport through the device (see Fig~\ref{fig:device}).  We calculate conductance through such a system as a function of temperature and gate voltage applied to the superconducting island. We show that transport is strongly affected by the presence of parafermionic zero modes, leading at zero temperature to a zero-bias conductance quantized in units of $\nu e^2/h$ independent of the applied gate voltage.
\end{abstract}

\maketitle

{\it Introduction.}
Topological superconductors, characterized by the presence of localized Majorana zero-energy modes (MZMs), have recently generated significant excitement in the condensed matter and quantum information communities~\cite{Nayak08, Beenakker13a, Alicea12a, Leijnse12, DasSarma15}.
Much of this excitement is due to the prediction that MZMs obey non-Abelian braiding statistics~\cite{Read00_Topological_SC_in_2D, Ivanov01, AliceaBraiding}, and as such have potential applications in topological quantum computation. Theory predicts that MZMs may be realized in semiconductor-superconductor heterostructures~\cite{Sau10,Alicea10,LutchynPRL10,1DwiresOreg}, and there is mounting experimental evidence for their existence in semiconductor nanowires~\cite{Mourik2012,Das2012,Deng2012,Fink2012,Churchill2013, Deng2014,Higginbotham15,albrecht2015,HaoZhang16, Deng16}. More recently, a number of proposals~\cite{Hyart13, Clarke16, Aasen16, Landau16,Plugge16a, Plugge16, Karzig16} were put forward describing how to realize a scalable platform for topological quantum computation using mesoscopic superconducting islands hosting two or more MZMs. The interplay between charging energy in mesoscale islands and topological degrees of freedom is an outstanding open problem.
	
In a normal-superconductor-normal (N-S-N) junction consisting of a gated s-wave superconducting island, the conductance through the device has $2e$-periodicity with the gate charge~\cite{Glazman'94, vonDelft'01}. The transport is dominated by the coherent Cooper-pair transmission through the island.
In contrast, an N-TSC-N junction has $e$-periodicity due to the presence of MZMs~\cite{Fu2010, vanHeck16, Lutchyn2017} which enable coherent single-electron transmission between opposite ends of a nanowire (i.e. an electron propagates coherently over distances much larger than the superconducting correlation length). This effect is at the heart of some of the recent measurement-only quantum computation proposals with Majorana zero modes~\cite{Plugge16, Karzig16}. An interesting question is whether this coherent transmission phenomenon has some analogue in \emph{fractional} 1D topological superconductors (fTSCs).

\begin{figure}
	\includegraphics[width=8cm]{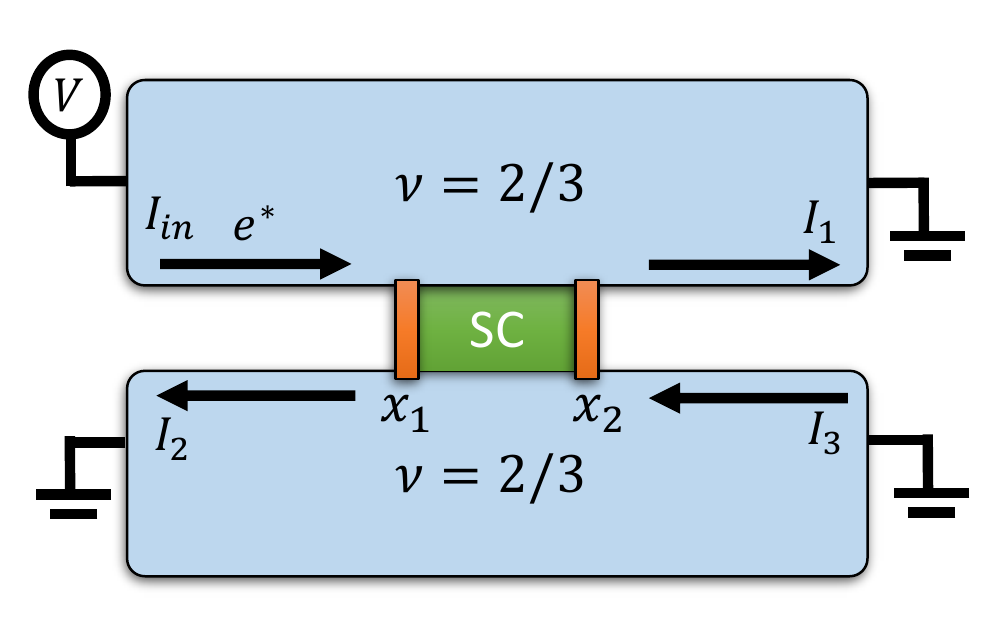}
	\caption{Schematic diagram of the device layout. Two counterpropagating fQH edges with the filling fraction $\nu=2/3$ are coupled to a floating mesoscopic s-wave superconductor with a charging energy $E_C$. In the appropriate parameter regime (see below), two parafermionic zero modes are localized at $x_1$ and $x_2$. Here the distance $|x_2-x_1|$ should be much larger than superconducting coherence length. The low-energy transport is dominated by the coherent charge $e^*$ transmission through the superconducting island.}
	\label{fig:device}
\end{figure}

One-dimensional (1D) fTSCs are characterized by the modes at their endpoints that may accommodate a discrete fraction of an electron charge $e^*$ at no energy cost. These modes, known as parafermionic zero modes, are a generalization of the more well-known Majorana zero modes, which can accommodate only electrons at no cost. According to a classification theorem~\cite{Fidkowski11a} parafermionic zero modes are forbidden in a generic purely one-dimensional system. However, 1D fTSCs may exist in \emph{effectively} 1D systems that emerge at the boundary of a 2D region that already admits fractionalized excitations, such as a 2D electron gas in a fractional quantum Hall (fQH) state. There have been several proposals for realizing these fractional topological superconductors in solid-state systems~\cite{Clarke2013a, Lindner2012, Cheng2012, Barkeshli16}. Recently, Clarke {\it et al.}~\cite{Clarke13} argued that fTSCs may lead to an interesting and unique set of circuit elements when the proximitizing superconductor is grounded (i.e. has no charging energy). In this Letter we consider a device (shown in Fig.\ref{fig:device}) with a floating fTSC and investigate the effect of charging energy on transport in such a system. We find that the transport properties of an fTSC in the presence of charging energy are drastically different from that in Majorana islands~\cite{albrecht2015,Fu2010,Zazunov11, vanHeck16, Lutchyn16, Lutchyn2017}. Floating metallic islands coupled to QH edges have been already realized experimentally~\cite{Iftikhar'15, Jezouin'16}. Therefore, we believe that our proposal is within the experimental reach, and is particularly suitable for graphene-based fTSC proposals.
	
{\it Theoretical model.}
    We consider the transport through a mesoscopic superconducting island connecting two counterpropagating fQH states at a filling fraction $\nu=2/3$ (see Fig.\ref{fig:device}).  We assume that edge states are strongly coupled to the superconductor in the region $x_1<x<x_2$, and are completely decoupled outside. Each edge state can be described using the $K$-matrix formalism~\cite{Wen1992a} with the corresponding Lagrangian $\mathcal{L}^{R/L}_0=\mathcal{L}^{R/L}_\rho+\mathcal{L}^{R/L}_\sigma$, where
    \begin{eqnarray}
     \mathcal{L}^{R/L}_\rho&=& \frac{3}{8\pi}\partial_x\phi_\rho\left(\pm
    \partial_{t}\phi_\rho-V_{\rho}\partial_x\phi_\rho\right)\label{eq:Lrho}\\
     \mathcal{L}^{R/L}_\sigma&=& -\frac{1}{8\pi}\partial_x\phi_\sigma
    \left(\pm\partial_{t}\phi_\sigma+V_{\sigma}\partial_x\phi_\sigma\right)\label{eq:Lsigma}.
    \end{eqnarray}
    Here $R/L$ denotes the right/left propagating edge modes. The fields $\phi_\rho$ and $\phi_\sigma$ correspond to charge and neutral modes, respectively. At $\nu=2/3$, the operator $e^{i\phi_\rho}$ creates a quasiparticle with charge $e^*=\nu e$. Note that we assume here that the $\nu=2/3$ state is unpolarized and neglect  spin-$SU(2)$ symmetry-breaking terms such as $V_{\rho\sigma}\partial_x\phi_\rho\partial_x\phi_\sigma$.

In terms of the chiral fields, electron operators at either side of the superconducting island can be written as
\ba
\psi^{R/L}_\up&=&\frac{1}{\sqrt{2\pi a}}e^{i(\frac32\phi^{R/L}_\rho-\frac12\phi^{R/L}_\sigma)},\\
\psi^{R/L}_\dn&=&\frac{1}{\sqrt{2\pi a}}e^{i(\frac32\phi^{R/L}_\rho+\frac12\phi^{R/L}_\sigma)},
\ea
where $a$ is a short-distance cutoff. We now introduce the non-chiral bosonic variables
$\phi^R_{\rho,\sigma}=\varphi_{\rho,\sigma}+\theta_{\rho,\sigma}$ and $\phi^L_{\rho,\sigma}=\varphi_{\rho,\sigma}-\theta_{\rho,\sigma}$ where the charge and spin fields satisfy the following commutation relations:
\ba
\left[\theta_\rho(x),\varphi_\rho(x')\right]&=&-\frac{2\pi i}{3}\Theta(x'-x),\\
\left[\theta_\sigma(x),\varphi_\sigma(x')\right]&=&2\pi i\Theta(x'-x).
\ea
Here $\Theta(x'-x)$ is the Heaviside theta function. The total charge density now reads
\be
\rho=\frac{\partial_x(\phi^R_\rho-\phi^L_\rho)}{2\pi}=\frac{\partial_x\theta_\rho}{\pi},
\ee
and the current operator for the corresponding segment of the four-terminal device shown in Fig.~\ref{fig:device} is given by \footnote{Here we have chosen a gauge in which all the currents in the system are flowing along the boundary of the quantum Hall regions that lies between the superconductor and the four leads, while none flows between the leads along the uppermost or lowermost edge in Fig.~\ref{fig:device}}
\begin{align}\label{eq:current}
I_{in}&=\frac{e}{2\pi}\dot{\phi}_\rho^R(x_1),\hspace{1.4cm}  I_1=\frac{e}{2\pi}\dot{\phi}_\rho^R(x_2),\\
I_2&=\frac{e}{2\pi}\dot{\phi}_\rho^L(x_1), \hspace{1.4cm} I_3=\frac{e}{2\pi}\dot{\phi}_\rho^L(x_2).
\end{align}
 The injected current in the linear response regime is given by $\langle I_{in}\rangle =V\nu e^2/h$ whereas the injected current $\langle I_3 \rangle=0$ since both contacts upstream of the bottom right edge are grounded. Therefore, we can define differential conductances for the two different drain electrodes $G_1=d\langle I_1 \rangle/dV$ and $G_2=d\langle I_2 \rangle /dV$ with the constraint $G_1+G_2=\nu e^2/h$ due to current conservation. After including interaction terms across the superconducting island
\be
\mathcal{L}^{RL}_C=-\frac{3}{8\pi} V^C_\rho \partial_x\phi^R_\rho \partial_x\phi^L_\rho-\frac{1}{8\pi} V^C_\sigma \partial_x\phi^R_\sigma \partial_x\phi^L_\sigma,
\ee
one arrives at the effective action $S_{\rho}+S_{\sigma}$ with
\ba
S_{\rho}&=&\frac{1}{2\pi}\int dxd\tau \frac{3K_\rho}{2}\left[ \frac{(\partial_\tau\varphi_\rho)^2}{v}+v(\partial_x\varphi_\rho)^2\right], \\
S_{\sigma}&=&\frac{1}{2\pi}\int dxd\tau \frac{K_\sigma}{2}\left[ \frac{(\partial_\tau\varphi_\sigma)^2}{v}+v(\partial_x\varphi_\sigma)^2\right]
\ea
where
$
K_{\rho,\sigma}=\sqrt{\frac{V_{\rho,\sigma}+V^C_{\rho,\sigma}}{V_{\rho,\sigma}-V^C_{\rho,\sigma}}}.
$
From now on we will assume a weak repulsive interaction between the charge modes and an attractive interaction between the neutral modes such that $K_\rho\gtrsim 1>K_\sigma$.

Next we consider various perturbations induced by the superconducting trench (of width smaller than the SC coherence length). In the Appendix \ref{app:A}, we analyze the single-particle and two-particle processes across the superconducting trench and calculate the scaling dimension of the corresponding operators. One can show that the neutral mode is gapped out in the singlet channel for $K_\sigma<1$ and, as a result, $\theta_\sigma$ is pinned.  Henceforth, we will assume that the gap for the neutral modes is the largest energy scale in the problem which effectively makes the system spinless. In the charge sector, there are two relevant bulk perturbations: a spin-conserving backscattering process $\psi^{\dagger R}_\up\psi^{L}_\up+\psi^{\dagger R}_\dn\psi^{L}_\dn+h.c.
\propto \cos(3\theta_\rho)$ and a superconducting pairing term in the singlet channel $\psi^{R}_\up\psi^{L}_\dn-\psi^{R}_\dn\psi^{L}_\up+h.c.
\propto \cos(3\varphi_\rho)$.
Both terms are relevant at $K_\rho\sim1$ and flow to strong coupling. However, given that $\theta_\rho$ and $\varphi_\rho$ are dual variables, these terms compete with each other and cannot order simultaneously. Henceforth, we focus on the limit when superconducting pairing dominates over backscattering term and opens a pairing gap in the trench, see detailed discussion in Refs.~\cite{Clarke2013a, Lindner2012, Cheng2012}. As a result, the backscattering term is suppressed in the bulk but may be important at the boundaries of the superconducting region ($x_1$, $x_2$). Note that the system with a grounded superconductor was   considered in Ref.~\cite{Clarke13} where it was shown that the parafermionic zero modes emerging at the end of the superconductor lead to a spectral flow of the boundary conditions and strongly modify transport properties of the system. In the present case, we consider a floating superconducting island with a finite charging energy $E_C \gg T$ with $T$ being the temperature. Thus, in contrast with Ref.~\cite{Clarke13}, uncorrelated Andreev processes at $x_1$ and $x_2$ are suppressed in our case.

Taking into account the above considerations, one can now write an effective low-energy model for the system. In the limit of weak backscattering at $x_{1/2}$, the corresponding Hamiltonian becomes
\begin{equation}
H=H_0+H_{B}+H_{P}+H_C
\end{equation}
where $H_0$ describes the two decoupled edges and
\begin{eqnarray}
H_{B}&=&-Dr_1 \cos(3\theta_\rho(x_1))-Dr_2\cos(3\theta_\rho(x_2)),\\
H_{P}&=&-\frac{\Delta}{2\pi a}\int_{x_1^+}^{x_2^-}dx \cos(3\varphi_\rho(x)),\\
H_C&=&E_C\left(\frac{\theta_\rho(x_2)-\theta_\rho(x_1)}{\pi}-\mathcal{N}_g\right)^2.
\end{eqnarray}
Here $r_{1,2} \ll 1$ are the reflection amplitudes at $x=x_{1,2}$, respectively, $\Delta$ is the induced SC gap, $E_C$ is the charging energy determined by the geometric capacitance of the island, $a$ is  the short-distance cutoff, and $x_i^{\pm}\equiv x_i \pm 0^+$. The charge on the island, given by $[\theta_\rho(x_2)-\theta_\rho(x_1)]/\pi$, can be tuned with the dimensionless gate voltage $\mathcal{N}_g=C_g V_g$ where $C_g$ and $V_g$ are gate capacitance and voltage, respectively. We implicitly assume here that due to the presence of a metallic island and strong hybridization between edge states and states in the metal, normal-state level spacing in the domain $x \in (x_1, x_2)$ becomes negligibly small.

{\it High-temperature limit}. We first analyze the high-temperature limit $E_C \gg T\gg \Delta$ when the island is in the normal state. At energies below $E_C$, charge fluctuations will be suppressed, resulting in the constraint $\theta^-_\rho\equiv  \theta_{\rho}(x_2)-\theta_{\rho}(x_1)=\pi {\mathcal N}_g$.  In terms of the fluctuating field $\theta_{\rho}^+\equiv\theta_{\rho}(x_2)+\theta_{\rho}(x_1)$, the boundary backscattering Hamiltonian $H_B$ is given by
    \begin{equation}
        H^{(\mathrm{eff})}_B=-Dr(\mathcal {N}_g)\cos\left(\frac{3}{2}\theta_{\rho}^+ -\beta(\mathcal{N}_g)\right)\label{eq:HBeff},
    \end{equation}
where $\beta(\mathcal{N}_g)$ is some unimportant phase, and $r(\mathcal{N}_g)$ reads
    \begin{align}
      \!\!r(\mathcal{N}_g)\!=\!\sgn\!\left(\!\cos{\frac{3\pi \mathcal{N}_g}{2}}\!\right)\!\!\sqrt{\!r_1^2\!+\!r_2^2\!+\!2r_1r_2\!\cos\left(3\pi \mathcal{N}_g\right)}.\label{eq:rNG}
    \end{align}
As a result of pinning of $\theta^-_\rho$, the RG equation for $r(\mathcal{N}_g)$ in the case $D<E_C$  becomes
$dr/dl=\left(1-3K_\rho/4\right)r$.
For $K_{\rho}<4/3$, the backscattering term is relevant, and flows to the strong coupling limit with $\theta_{\rho}^+$ pinned. Using the condition $r(D_c)\sim 1$, we find the strong-coupling crossover scale $D_c$:
\begin{equation}
 D_c\sim E_C r(\mathcal{N}_g)^\frac{4}{4-3K_{\rho}}.
\end{equation}
In the intermediate regime $D_c \ll D \ll E_C$, the backscattering term remains small and can be taken into account perturbatively.

The differential tunneling conductance in different temperature regimes can be evaluated using the Kubo formula~\cite{Aleiner'02}
\begin{align}\label{eq:kubo}
G_i\!=\!\frac{1}{2T}\int_{-\infty}^{\infty}\!dt {\Pi}_i\left(it +\frac{1}{2T}\right), \,  \Pi_i (\tau)\!=\!\langle I_i(\tau)I_i(0) \rangle
\end{align}
Here $\tau$ is imaginary time, and $I_i$ is the corresponding expression for the current operator, see Eq.~\eqref{eq:current}. The resulting conductance $G_1(T)$ for ${\rm max} \{D_c, \Delta \} \ll T \ll E_C$ is given by
\be \label{eq:G1normal}
\frac{G_1(T)}{G_0}=\nu \left(1- c_1 r(\mathcal{N}_g)^2\left(\frac{E_C}{T}\right)^{\frac{4-3K_\rho}{2}}\right),
\ee
where $G_0=e^2/h$ and $c_1$ is an $O(1)$ numerical constant.

Let us now consider the case $\Delta \ll T \ll D_c$ ~\footnote{In Appendix B, we present the cases $E_C\gg\Delta\gg D_c$ and $\Delta \gg E_C$. Using similar RG analysis, one may show that in both these cases $G_1(T)=\nu e^2/h$ with a temperature-dependent correction scaling as $T^{3K_\rho-2}$ for $T\ll {\rm min} \{E_C, \Delta \}$.} where backscattering becomes large and the system flows to strong coupling, thereby pinning the field $\theta_\rho^+$ at the boundary.
In order to calculate the conductance in this case, we first need to perform a duality transformation. The leading irrelevant operator, which shifts $\frac{3}{2} \theta_\rho^+$ by $2\pi$, is given by
\begin{equation}\label{eq:cotunneling}
H_{\mathrm{dual}}=-D\lambda(D) \cos\left(\delta\varphi_{\mathrm{out}}-\delta\varphi_{\mathrm{in}}\right),
\end{equation}
where $\delta\varphi_{\mathrm{out}}=\varphi_{\rho}(x_2^+)-\varphi_{\rho}(x_1^-)$ and $\delta\varphi_{\mathrm{in}}=\varphi_{\rho}(x_2^-)-\varphi_{\rho}(x_1^+)$. Eq.\eqref{eq:cotunneling} describes a process of correlated tunneling of charge $e^*$ at $x_1$ and $x_2$ preserving the total charge in the island. The scaling dimension of this operator is $4/3K_{\rho}$, in keeping with its role as the dual of the Hamiltonian $H^{(\mathrm{eff})}_B$. The RG flow for $\lambda$ reads $d\lambda/dl=\left(1-4/3K_\rho\right)\lambda$.
Let us now consider transport at this fixed point. The pinning of boundary fields $\theta_\rho^{\pm}$ implies that
\be
\dot{\phi}_\rho^R(x_{1})=\dot{\phi}^L_\rho(x_{1}) \text{ and } \dot{\phi}_\rho^R(x_{2})=\dot{\phi}^L_\rho(x_{2}).
\ee
Thus, there is strong backscattering at $x_1$ and $x_2$ resulting in $\langle I_1 \rangle \rightarrow 0$. Assuming $\Delta \ll T \ll D_c$, the conductance $G_1(T)$ can be calculated perturbatively in $\lambda$. Using the Kubo formula~\eqref{eq:kubo} and the current-conservation constraint at $x_1$ (i.e. $G_2=\nu G_0-G_1$), one finds that
\be
\frac{G_{1}(T)}{G_0} \sim \lambda(D_c)^2 \left(\frac{T}{D_c}\right)^{\frac{8}{3K_\rho}\!-\!2}\sim r(\mathcal{N}_g)^{-\frac{8}{3K_\rho}} \left(\frac{T}{E_C}\right)^{\frac{8}{3K_\rho}\!-\!2}\label{eq:G1dual},
\ee
where we used $\lambda(D_c) \sim 1$. Thus, transport through the island in this temperature regime is dominated by the inelastic processes and is suppressed at low temperatures.

{\it Low-temperature limit}. Let us now consider the low-temperature limit $T \ll \Delta $. We expect that transport properties will be significantly modified due to presence of parafermionic zero modes~\cite{Clarke2013a, Lindner2012, Cheng2012, Clarke13}. In the limit $\Delta \ll D_c$, the effective Hamiltonian at the scale $D \sim D_c$ is given by Eq.~\eqref{eq:cotunneling} with $\lambda(D)=\lambda(D_c) (D/D_c)^{4/3K_{\rho}-1}$.
Upon lowering the bandwidth to $D\sim \Delta$, the SC pairing $H_P$ opens a gap in the spectrum and suppresses fluctuations of $\delta \varphi_{in}$.
It is illuminating to rewrite the low-energy boundary Hamiltonian~\eqref{eq:cotunneling} in terms of the parafermionic zero modes. Using the right-moving representation~\footnote{right- and left-moving representations are not independent, and one can equivalently write the Hamiltonian in terms of left-movers\cite{Clarke2013a,Mong14a,Mong14b}.}, the effective  Hamiltonian becomes
\begin{equation}
H_{\mathrm{dual}}\!=\!-\frac12 D\lambda^*(D) e^{i\phi^R_\rho(x_2^+)}e^{-i\phi^R_\rho(x_1^-)}\alpha^{R\dagger}_2\alpha_1^R+\mathrm{h.c.}, \label{eq:dualpara}
\end{equation}
where $\alpha_{1,2}^{R}$ are parafermionic operators localized at $x_{1/2}$. One should keep in mind that the system hosting two parafermionic zero modes ($N_m=2$) does not have ground-state degeneracy since charge on the island is fixed by the charging energy. If, however, the number of zero modes $N_m > 2$, ground-state degeneracy will be restored and the process considered above provides a way of measuring which ground-state the system is in. Hamiltonian~\eqref{eq:dualpara} describes a coherent transfer of charge $e^*$ quasiparticles through the superconducting island, and is reminiscent of the single-electron coherent transmission in Majorana systems~\cite{Fu2010, Lutchyn2017}.

Let's now analyze transport properties  at low temperature $T \ll \Delta$. One may notice that the scaling dimension of $\lambda(D)$ for $D < \Delta$  is halved to $2/3K_{\rho}$.  Thus, the boundary term~\eqref{eq:dualpara} becomes relevant for $2/3<K_{\rho}$, and $\lambda (D)$ grows under RG and reaches strong coupling limit at the new scale:
\be
D_s\sim \Delta \cdot \lambda(\Delta)^{\frac{3K_\rho}{3K_\rho-2}}.
\ee
Using $\lambda(\Delta)\sim (\Delta/D_c)^{\frac{4}{3K_\rho}-1}$, the differential conductance can be calculated perturbatively in the limit $D_s\ll T \ll \Delta$ yielding
\be
\frac{G_1(T)}{G_0}\sim r(\mathcal{N}_g)^{-\frac{8}{3K_\rho}} \left(\frac{\Delta}{E_C}\right)^{\frac{8}{3K_\rho}-2}\left(\frac{\Delta}{T}\right)^{2-\frac{4}{3K_\rho}}.
\ee
Notice that above expression matches Eq. \eqref{eq:G1dual} at $T\sim \Delta$.
\begin{figure}[t]
	\includegraphics[width=8cm]{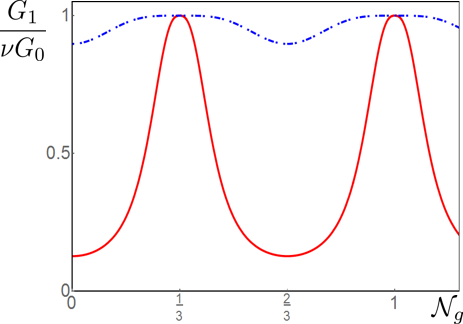}
	\caption{(Color online) Schematic plot of the differential conductance $G_1(\mathcal{N}_g)$ in the low-temperature $T\ll  D_s, \Delta$ (dash-dot blue line) and high-temperature $\Delta \ll T \ll D_c, E_C$ (solid red line) limits.  Here we assume symmetric contacts $r_1=r_2$.}\label{fig:conductance}
\end{figure}

Finally, let's consider the low-temperature regime $T\ll D_s, \Delta$. At $D < D_s$, the boundary condition for the fields becomes $\varphi_{\rho}(x_2^+)-\varphi_{\rho}(x_1^-)=\rm const$ which leads to the following conservation law for the chiral fields
\be
\dot{\phi}^R_\rho(x_2)+\dot{\phi}^L_\rho(x_2)-\dot{\phi}^R_\rho(x_1)-\dot{\phi}^L_\rho(x_1)=0.
\ee
Using current conservation, one finds that $\langle I_1 \rangle = \langle I \rangle $ and $\langle I_2 \rangle =0$. As a result, we conclude that zero-temperature conductance $G_{1}=\nu e^2/h$ and is independent of $\mathcal{N}_g$ which is very different from the Majorana case~\cite{Lutchyn2017}. Finite-temperature corrections to the conductance can be calculated by perturbing the above result with the leading irrelevant operator at the strong coupling fixed point $D \ll D_s$:
    \begin{equation}\label{eq:effBM}
      H^\mathrm{eff}_B\sim- D r(D) \cos\left(\frac{3\theta_\rho^+}{2}\right).
    \end{equation}
Given that $\varphi_\rho$ is pinned in the domain $x_1<x<x_2$, the RG flow for $r(D)$ becomes $dr/dl= (1- 3 K_\rho /2) r$.
Thus, at the energy scale $D \ll D_s$, one finds that
    \begin{equation}
     \tilde{r}(D)=\tilde{r}(D_s)\left(\frac{D}{D_s}\right)^{\frac{3K_\rho}{2}-1}.
    \end{equation}
By perturbatively evaluating corrections to the conductance using the Kubo formula ~\eqref{eq:kubo}  (see Ref.\cite{Lutchyn2013} for details) one finds
		\be
		\frac{G_{1}(T)}{G_0}\!=\! \nu \left(1-c_2 r^4(\mathcal{N}_g) \left(\frac{E_C}{\Delta}\right)^{4-3K_\rho}\!\left(\frac{T}{\Delta}\right)^{3K_\rho-2}\right)\label{eq:G}.
		\ee	
Here $c_2$ is an $O(1)$ numerical coefficient.  This is a counterintuitive result. Despite the fact that the backscattering term $D_c$ was initially large ({\it i.e.} $D_c \gg \Delta$), the low-energy transport properties are characterized by a universal value of the conductance. In other words, ground-state properties of the system are independent of  $\mathcal{N}_g$ (i.e.  effective charging energy is renormalized to zero by quantum fluctuations).

Let's compare our results for the Coulomb blockade in the fractional TSC systems with the corresponding case in the Majorana counterparts~\cite{Fu2010, Zazunov11,vanHeck16, Lutchyn16, Lutchyn2017}. In the Majorana systems
the backscattering operator is marginal~\cite{Lutchyn16} and the zero-temperature conductance $G_1$ is dependent on $\mathcal{N}_g$: it reaches maximum of the order of $e^2/h$ at the charge degeneracy points and gets significantly reduced in the Coulomb valleys. In stark contrast, we find \emph{quantized} conductance $G_1$ in the fractional TSC systems. This drastic difference originates from the fact that backscattering operators for charge-$e^*$ quasiparticles are not allowed between fractional QH edges separated by the trivial vacuum and backscattering is therefore dominated by fermionic processes having higher scaling dimension. As a result, quantum charge fluctuations are much stronger in fTSC systems than in Majorana systems.

{\it Conclusion.}
Coulomb blockade of charge transport across a mesoscopic superconducting island manifests itself through the oscillations of the conductance with the gate voltage ${\cal N}_g$. In Majorana islands the periodicity of the oscillations corresponds to an increment of charge by $e$ whereas in fractional topological superconductors this periodicity is determined by the fractional quasiparticle charge $e^*$. In this Letter we have developed a framework for studying the Coulomb blockade effect in QH-superconductor heterostructures. By considering the specific fractional topological superconductor proposal based on $\nu=2/3$ QH state, we show that dependence of the differential conductance on gate voltage and temperature is quite non-trivial. At zero temperature the conductance approaches a quantized value of $\nu e^2/h$. The dependence on gate voltage appears only at finite temperature with the amplitude of gate-voltage oscillations increasing with temperature (see Eq. \eqref{eq:G}). The conductance decreases with increasing temperature until $T$ reaches the superconducting gap scale $\Delta$ and then increases again to the quantized value for $\Delta \ll T\ll E_C$.

This work was supported by  LPS-MPO-CMTC, JQI-NSF-PFC (DJC) and Microsoft (YK and DJC). We acknowledge stimulating discussions with P. Bonderson, T. Karzig, D. E. Liu,  C. Nayak.
and D. Pikulin.

\appendix

\section{Analysis of the bulk perturbations across the trench}\label{app:A}

In this Appendix, we analyze different perturbations across the superconducting trench. As shown below, there are six terms which one can write using different combinations of electrons from the two edges, each of which is marginal in the absence of interactions.
\begin{widetext}
\begin{itemize}
	\item Spin-conserving backscattering
	\ba
	\mathcal{O}_{sb} = \Delta_{sb} (\psi^{\dagger R}_\up\psi^{L}_\up+\psi^{\dagger R}_\dn\psi^{L}_\dn)+h.c. \propto \cos(3\theta_\rho-\alpha_{sh})\cos(\theta_\sigma-\beta_{sh})
	\ea
	has scaling dimension $\frac12 K_\sigma+\frac32 K_\rho$.
	\item Spin-flip backscattering
	\ba
	\mathcal{O}_{tb} = \Delta_{tb} (\psi^{\dagger R}_\up\psi^{L}_\dn-\psi^{\dagger R}_\dn\psi^{L}_\up)+h.c. \propto \cos(3\theta_\rho-\alpha_{th})\cos(\varphi_\sigma-\beta_{th})
	\ea
	has scaling dimension $\frac12 K_\sigma^{-1}+\frac32 K_\rho$.
	\item singlet pairing
	\ba
	\mathcal{O}_{sp} = \Delta_{sp} (\psi^{R}_\up\psi^{L}_\dn-\psi^{R}_\dn\psi^{L}_\up)+h.c. \propto \cos(3\varphi_\rho-\alpha_{sp})\cos(\theta_\sigma-\beta_{sp})
	\ea
	has scaling dimension $\frac12 K_\sigma+\frac32 K_\rho^{-1}$.
	\item triplet pairing
	\ba
	\mathcal{O}_{tp} = \Delta_{tb} (\psi^{R}_\up\psi^{L}_\up+\psi^{ R}_\dn\psi^{L}_\dn)+h.c. \propto \cos(3\varphi_\rho-\alpha_{tp})\cos(\varphi_\sigma-\beta_{tp})
	\ea
	has scaling dimension $\frac12 K_\sigma^{-1}+\frac32 K_\rho^{-1}$.
	\item neutral singlet four-fermion coupling
	\ba
	\mathcal{O}_{ns} = \Delta_{ns} \psi^{\dagger R}_\up\psi^{R}_\dn\psi^{\dagger L}_\dn\psi^{L}_\up+h.c. \propto \cos(2\theta_\sigma-\beta_{ns})
	\ea
	has scaling dimension $2 K_\sigma$.
	\item neutral triplet four-fermion coupling
	\ba
	\mathcal{O}_{nt} = \Delta_{nt} \psi^{\dagger R}_\up\psi^{R}_\dn\psi^{\dagger L}_\up\psi^{L}_\dn+h.c. \propto \cos(2\varphi_\sigma-\beta_{nt})
	\ea
	has scaling dimension $2 K_\sigma^{-1}$.
\end{itemize}
\end{widetext}
Here we assume that $SU(2)$ spin symmetry is preserved and the edges are equivalent.

Now we consider a general model Hamiltonian where we induce the coupling between the two sides of the superconducting trench
\be
H=H_0+ \int_{x_1}^{x_2}dx \delta H
\ee
where $H_0$ is the bulk Hamiltonian and
$\delta H$ contains all the perturbations $\delta H= \sum_{i} \mathcal{O}_i$.

The corresponding lowest-order RG equations for the aforementioned operators are
\ba
\frac{d\Delta_{sb}}{dl}&=&\left(2-\frac{K_\sigma}{2}-\frac{3K_\rho}{2}\right)\Delta_{sb},\\
\frac{d\Delta_{tb}}{dl}&=&\left(2-\frac{1}{2K_\sigma}-\frac{3K_\rho}{2}\right)\Delta_{tb},\\
\frac{d\Delta_{sp}}{dl}&=&\left(2-\frac{K_\sigma}{2}-\frac{3}{2K_\rho}\right)\Delta_{sp},\\
\frac{d\Delta_{tp}}{dl}&=&\left(2-\frac{1}{2K_\sigma}-\frac{3}{2K_\rho}\right)\Delta_{tp},\\
\frac{d\Delta_{ns}}{dl}&=&(2-2K_\sigma)\Delta_{ns},\\
\frac{d\Delta_{nt}}{dl}&=&\left(2-\frac{2}{K_\sigma}\right)\Delta_{nt}.
\ea
One may notice that for $K_\sigma<1$, $\Delta_{sb}$, $\Delta_{sp}$ and $\Delta_{ns}$ are relevant perturbations in the bulk and flow to strong coupling. Neutral modes are gapped by the $\Delta_{ns}$ term independently of the value for $K_\rho$ and, thus, one may ignore them. For $K_\rho \approx 1$, both couplings $\Delta_{sb}$ and $\Delta_{sp}$ are relevant and compete with each other. In this paper we assume that initial values $\Delta_{sb}(l_0) \ll \Delta_{sp}(l_0)$ so that the singlet pairing term $\Delta_{sp}$ dominates and reaches strong coupling limit first. 

\section{Conductance calculation for a large superconducting gap}

In this section, we present the calculation of the conductance in a Coulomb blockade regime (i.e. $T \ll E_C$) in two different parameter regimes:  a) $E_C\gg \Delta(\gg D_c)\gg T$ and b) $\Delta  \gg E_C \gg T$. We show below that the low-temperature conductance is approaching $\nu e^2/h$ with temperature corrections scaling as $T^{3K_\rho-2}$. The difference with respect to Eq. (31) of the main text appears in the prefactor of the temperature-dependent correction. 

First, let us consider the limit when $E_C \gg \Delta \gg D_c$. In this case, the induced superconducting pairing opens a gap in the spectrum at $D\sim \Delta \gg D_c$. As a result, the RG flow of the backscattering term is modified. At $D < \Delta$, $\varphi_\rho$ is pinned inside the region $x_1<x<x_2$, and the RG equation for backscattering $r(\mathcal{N}_g)$ reads
\be
\frac{dr}{dl}=\left(1-\frac{3K_\rho}{2}\right)r \label{eq:rgflow}
\ee
Thus, the backscattering process becomes irrelevant now for $K_\rho>2/3$, and $r$ does not reach the strong coupling limit. The conductance at $T\ll \Delta$ can be calculated perturbatively in $r$ and is given by
\begin{align}\label{eq:G11}
\frac{G_{1}(T)}{G_0}=\nu\left(1-r^2(\mathcal{N}_g)  \left(\frac{E_C}{\Delta}\right)^{2-\frac{3}{2}K_\rho}\left(\frac{T}{\Delta}\right)^{3K_\rho-2}\right).
\end{align}
Notice that above exprresion for the conductance has the same temperature dependence as in Eq. (31) of the main text and matches it at $\Delta=D_c$.  

Next, we consider the second limit when $\Delta \gg E_C$. At the bandwidth $D=E_c$, we once again pin the combination $\theta_\rho(x_2)-\theta_\rho(x_1)$ and define $r(\mathcal{N}_g)$. Since the field $\varphi_\rho$ inside the superconducting island is already pinned by $\Delta$, the effective backscattering amplitude flows according to Eq.\eqref{eq:rgflow}, and is irrelevant for $K_\rho>2/3$. The conductance can be calculated perturbatively in $r$ and for $T\ll E_C\ll \Delta$ is given by
\begin{align}\label{eq:G111}
\frac{G_{1}(T)}{G_0}=\nu\left(1-r^2(\mathcal{N}_g)  \left(\frac{T}{E_C}\right)^{3K_\rho-2}\right).
\end{align}
As in the previous case where $E_C$ was the largest energy scale, we find that the system shows perfect transmission across the superconducting region at zero temperature with the same power $3K_\rho-2$ for the temperature-dependent correction. 
Note that Eqs.\eqref{eq:G11} and \eqref{eq:G111} match at $E_C\sim \Delta$.

\bibliography{Fcb}

\begin{thebibliography}{52}%
\makeatletter
\providecommand \@ifxundefined [1]{%
 \@ifx{#1\undefined}
}%
\providecommand \@ifnum [1]{%
 \ifnum #1\expandafter \@firstoftwo
 \else \expandafter \@secondoftwo
 \fi
}%
\providecommand \@ifx [1]{%
 \ifx #1\expandafter \@firstoftwo
 \else \expandafter \@secondoftwo
 \fi
}%
\providecommand \natexlab [1]{#1}%
\providecommand \enquote  [1]{``#1''}%
\providecommand \bibnamefont  [1]{#1}%
\providecommand \bibfnamefont [1]{#1}%
\providecommand \citenamefont [1]{#1}%
\providecommand \href@noop [0]{\@secondoftwo}%
\providecommand \href [0]{\begingroup \@sanitize@url \@href}%
\providecommand \@href[1]{\@@startlink{#1}\@@href}%
\providecommand \@@href[1]{\endgroup#1\@@endlink}%
\providecommand \@sanitize@url [0]{\catcode `\\12\catcode `\$12\catcode
  `\&12\catcode `\#12\catcode `\^12\catcode `\_12\catcode `\%12\relax}%
\providecommand \@@startlink[1]{}%
\providecommand \@@endlink[0]{}%
\providecommand \url  [0]{\begingroup\@sanitize@url \@url }%
\providecommand \@url [1]{\endgroup\@href {#1}{\urlprefix }}%
\providecommand \urlprefix  [0]{URL }%
\providecommand \Eprint [0]{\href }%
\providecommand \doibase [0]{http://dx.doi.org/}%
\providecommand \selectlanguage [0]{\@gobble}%
\providecommand \bibinfo  [0]{\@secondoftwo}%
\providecommand \bibfield  [0]{\@secondoftwo}%
\providecommand \translation [1]{[#1]}%
\providecommand \BibitemOpen [0]{}%
\providecommand \bibitemStop [0]{}%
\providecommand \bibitemNoStop [0]{.\EOS\space}%
\providecommand \EOS [0]{\spacefactor3000\relax}%
\providecommand \BibitemShut  [1]{\csname bibitem#1\endcsname}%
\let\auto@bib@innerbib\@empty
\bibitem [{\citenamefont {Nayak}\ \emph {et~al.}(2008)\citenamefont {Nayak},
  \citenamefont {Simon}, \citenamefont {Stern}, \citenamefont {Freedman},\ and\
  \citenamefont {Sarma}}]{Nayak08}%
  \BibitemOpen
  \bibfield  {author} {\bibinfo {author} {\bibfnamefont {C.}~\bibnamefont
  {Nayak}}, \bibinfo {author} {\bibfnamefont {S.~H.}\ \bibnamefont {Simon}},
  \bibinfo {author} {\bibfnamefont {A.}~\bibnamefont {Stern}}, \bibinfo
  {author} {\bibfnamefont {M.}~\bibnamefont {Freedman}}, \ and\ \bibinfo
  {author} {\bibfnamefont {S.~D.}\ \bibnamefont {Sarma}},\ }\href@noop {}
  {\bibfield  {journal} {\bibinfo  {journal} {Rev. Mod. Phys.}\ }\textbf
  {\bibinfo {volume} {80}},\ \bibinfo {pages} {1083} (\bibinfo {year}
  {2008})}\BibitemShut {NoStop}%
\bibitem [{\citenamefont {Beenakker}(2013)}]{Beenakker13a}%
  \BibitemOpen
  \bibfield  {author} {\bibinfo {author} {\bibfnamefont {C.~W.~J.}\
  \bibnamefont {Beenakker}},\ }\href {\doibase
  10.1146/annurev-conmatphys-030212-184337} {\bibfield  {journal} {\bibinfo
  {journal} {Annu. Rev. Condens. Matter Phys.}\ }\textbf {\bibinfo {volume}
  {4}},\ \bibinfo {pages} {113} (\bibinfo {year} {2013})},\ \Eprint
  {http://arxiv.org/abs/arXiv:1112.1950} {arXiv:1112.1950} \BibitemShut
  {NoStop}%
\bibitem [{\citenamefont {Alicea}(2012)}]{Alicea12a}%
  \BibitemOpen
  \bibfield  {author} {\bibinfo {author} {\bibfnamefont {J.}~\bibnamefont
  {Alicea}},\ }\href@noop {} {\bibfield  {journal} {\bibinfo  {journal} {Rep.
  Prog. Phys.}\ }\textbf {\bibinfo {volume} {75}},\ \bibinfo {pages} {076501}
  (\bibinfo {year} {2012})},\ \Eprint {http://arxiv.org/abs/arXiv:1202.1293}
  {arXiv:1202.1293} \BibitemShut {NoStop}%
\bibitem [{\citenamefont {{Leijnse}}\ and\ \citenamefont
  {{Flensberg}}(2012)}]{Leijnse12}%
  \BibitemOpen
  \bibfield  {author} {\bibinfo {author} {\bibfnamefont {M.}~\bibnamefont
  {{Leijnse}}}\ and\ \bibinfo {author} {\bibfnamefont {K.}~\bibnamefont
  {{Flensberg}}},\ }\href {\doibase 10.1088/0268-1242/27/12/124003} {\bibfield
  {journal} {\bibinfo  {journal} {Semiconductor Science Technology}\ }\textbf
  {\bibinfo {volume} {27}},\ \bibinfo {eid} {124003} (\bibinfo {year}
  {2012})},\ \Eprint {http://arxiv.org/abs/1206.1736} {arXiv:1206.1736
  [cond-mat.mes-hall]} \BibitemShut {NoStop}%
\bibitem [{\citenamefont {{Das Sarma}}\ \emph {et~al.}(2015)\citenamefont {{Das
  Sarma}}, \citenamefont {{Freedman}},\ and\ \citenamefont
  {{Nayak}}}]{DasSarma15}%
  \BibitemOpen
  \bibfield  {author} {\bibinfo {author} {\bibfnamefont {S.}~\bibnamefont {{Das
  Sarma}}}, \bibinfo {author} {\bibfnamefont {M.}~\bibnamefont {{Freedman}}}, \
  and\ \bibinfo {author} {\bibfnamefont {C.}~\bibnamefont {{Nayak}}},\
  }\href@noop {} {\bibfield  {journal} {\bibinfo  {journal} {ArXiv e-prints}\ }
  (\bibinfo {year} {2015})},\ \Eprint {http://arxiv.org/abs/1501.02813}
  {arXiv:1501.02813 [cond-mat.str-el]} \BibitemShut {NoStop}%
\bibitem [{\citenamefont {Read}\ and\ \citenamefont
  {Green}(2000)}]{Read00_Topological_SC_in_2D}%
  \BibitemOpen
  \bibfield  {author} {\bibinfo {author} {\bibfnamefont {N.}~\bibnamefont
  {Read}}\ and\ \bibinfo {author} {\bibfnamefont {D.}~\bibnamefont {Green}},\
  }\href@noop {} {\bibfield  {journal} {\bibinfo  {journal} {Phys. Rev. B}\
  }\textbf {\bibinfo {volume} {61}},\ \bibinfo {pages} {10267} (\bibinfo {year}
  {2000})}\BibitemShut {NoStop}%
\bibitem [{\citenamefont {Ivanov}(2001)}]{Ivanov01}%
  \BibitemOpen
  \bibfield  {author} {\bibinfo {author} {\bibfnamefont {D.~A.}\ \bibnamefont
  {Ivanov}},\ }\href {\doibase 10.1103/PhysRevLett.86.268} {\bibfield
  {journal} {\bibinfo  {journal} {Phys. Rev. Lett.}\ }\textbf {\bibinfo
  {volume} {86}},\ \bibinfo {pages} {268} (\bibinfo {year} {2001})}\BibitemShut
  {NoStop}%
\bibitem [{\citenamefont {Alicea}\ \emph {et~al.}(2011)\citenamefont {Alicea},
  \citenamefont {Oreg}, \citenamefont {Refael}, \citenamefont {von Oppen},\
  and\ \citenamefont {Fisher}}]{AliceaBraiding}%
  \BibitemOpen
  \bibfield  {author} {\bibinfo {author} {\bibfnamefont {J.}~\bibnamefont
  {Alicea}}, \bibinfo {author} {\bibfnamefont {Y.}~\bibnamefont {Oreg}},
  \bibinfo {author} {\bibfnamefont {G.}~\bibnamefont {Refael}}, \bibinfo
  {author} {\bibfnamefont {F.}~\bibnamefont {von Oppen}}, \ and\ \bibinfo
  {author} {\bibfnamefont {M.~P.~A.}\ \bibnamefont {Fisher}},\ }\href {\doibase
  10.1038/nphys1915} {\bibfield  {journal} {\bibinfo  {journal} {Nature Phys.}\
  }\textbf {\bibinfo {volume} {7}},\ \bibinfo {pages} {412} (\bibinfo {year}
  {2011})}\BibitemShut {NoStop}%
\bibitem [{\citenamefont {Sau}\ \emph {et~al.}(2010)\citenamefont {Sau},
  \citenamefont {Lutchyn}, \citenamefont {Tewari},\ and\ \citenamefont
  {Das~Sarma}}]{Sau10}%
  \BibitemOpen
  \bibfield  {author} {\bibinfo {author} {\bibfnamefont {J.~D.}\ \bibnamefont
  {Sau}}, \bibinfo {author} {\bibfnamefont {R.~M.}\ \bibnamefont {Lutchyn}},
  \bibinfo {author} {\bibfnamefont {S.}~\bibnamefont {Tewari}}, \ and\ \bibinfo
  {author} {\bibfnamefont {S.}~\bibnamefont {Das~Sarma}},\ }\href@noop {}
  {\bibfield  {journal} {\bibinfo  {journal} {Phys. Rev. Lett.}\ }\textbf
  {\bibinfo {volume} {104}},\ \bibinfo {pages} {040502} (\bibinfo {year}
  {2010})}\BibitemShut {NoStop}%
\bibitem [{\citenamefont {Alicea}(2010)}]{Alicea10}%
  \BibitemOpen
  \bibfield  {author} {\bibinfo {author} {\bibfnamefont {J.}~\bibnamefont
  {Alicea}},\ }\href@noop {} {\bibfield  {journal} {\bibinfo  {journal} {Phys.
  Rev. B}\ }\textbf {\bibinfo {volume} {81}},\ \bibinfo {pages} {125318}
  (\bibinfo {year} {2010})}\BibitemShut {NoStop}%
\bibitem [{\citenamefont {Lutchyn}\ \emph {et~al.}(2010)\citenamefont
  {Lutchyn}, \citenamefont {Sau},\ and\ \citenamefont
  {Das~Sarma}}]{LutchynPRL10}%
  \BibitemOpen
  \bibfield  {author} {\bibinfo {author} {\bibfnamefont {R.~M.}\ \bibnamefont
  {Lutchyn}}, \bibinfo {author} {\bibfnamefont {J.~D.}\ \bibnamefont {Sau}}, \
  and\ \bibinfo {author} {\bibfnamefont {S.}~\bibnamefont {Das~Sarma}},\ }\href
  {\doibase 10.1103/PhysRevLett.105.077001} {\bibfield  {journal} {\bibinfo
  {journal} {Phys.\ Rev.\ Lett.}\ }\textbf {\bibinfo {volume} {105}},\ \bibinfo
  {pages} {077001} (\bibinfo {year} {2010})}\BibitemShut {NoStop}%
\bibitem [{\citenamefont {Oreg}\ \emph {et~al.}(2010)\citenamefont {Oreg},
  \citenamefont {Refael},\ and\ \citenamefont {von Oppen}}]{1DwiresOreg}%
  \BibitemOpen
  \bibfield  {author} {\bibinfo {author} {\bibfnamefont {Y.}~\bibnamefont
  {Oreg}}, \bibinfo {author} {\bibfnamefont {G.}~\bibnamefont {Refael}}, \ and\
  \bibinfo {author} {\bibfnamefont {F.}~\bibnamefont {von Oppen}},\ }\href
  {\doibase 10.1103/PhysRevLett.105.177002} {\bibfield  {journal} {\bibinfo
  {journal} {Phys.\ Rev.\ Lett.}\ }\textbf {\bibinfo {volume} {105}},\ \bibinfo
  {pages} {177002} (\bibinfo {year} {2010})}\BibitemShut {NoStop}%
\bibitem [{\citenamefont {{Mourik}}\ \emph {et~al.}(2012)\citenamefont
  {{Mourik}}, \citenamefont {{Zuo}}, \citenamefont {{Frolov}}, \citenamefont
  {{Plissard}}, \citenamefont {{Bakkers}},\ and\ \citenamefont
  {{Kouwenhoven}}}]{Mourik2012}%
  \BibitemOpen
  \bibfield  {author} {\bibinfo {author} {\bibfnamefont {V.}~\bibnamefont
  {{Mourik}}}, \bibinfo {author} {\bibfnamefont {K.}~\bibnamefont {{Zuo}}},
  \bibinfo {author} {\bibfnamefont {S.~M.}\ \bibnamefont {{Frolov}}}, \bibinfo
  {author} {\bibfnamefont {S.~R.}\ \bibnamefont {{Plissard}}}, \bibinfo
  {author} {\bibfnamefont {E.~P.~A.~M.}\ \bibnamefont {{Bakkers}}}, \ and\
  \bibinfo {author} {\bibfnamefont {L.~P.}\ \bibnamefont {{Kouwenhoven}}},\
  }\href {\doibase 10.1126/science.1222360} {\bibfield  {journal} {\bibinfo
  {journal} {Science}\ }\textbf {\bibinfo {volume} {336}},\ \bibinfo {pages}
  {1003} (\bibinfo {year} {2012})}\BibitemShut {NoStop}%
\bibitem [{\citenamefont {{Das}}\ \emph {et~al.}(2012)\citenamefont {{Das}},
  \citenamefont {{Ronen}}, \citenamefont {{Most}}, \citenamefont {{Oreg}},
  \citenamefont {{Heiblum}},\ and\ \citenamefont {{Shtrikman}}}]{Das2012}%
  \BibitemOpen
  \bibfield  {author} {\bibinfo {author} {\bibfnamefont {A.}~\bibnamefont
  {{Das}}}, \bibinfo {author} {\bibfnamefont {Y.}~\bibnamefont {{Ronen}}},
  \bibinfo {author} {\bibfnamefont {Y.}~\bibnamefont {{Most}}}, \bibinfo
  {author} {\bibfnamefont {Y.}~\bibnamefont {{Oreg}}}, \bibinfo {author}
  {\bibfnamefont {M.}~\bibnamefont {{Heiblum}}}, \ and\ \bibinfo {author}
  {\bibfnamefont {H.}~\bibnamefont {{Shtrikman}}},\ }\href {\doibase
  10.1038/nphys2479} {\bibfield  {journal} {\bibinfo  {journal} {Nature Phys.}\
  }\textbf {\bibinfo {volume} {8}},\ \bibinfo {pages} {887} (\bibinfo {year}
  {2012})}\BibitemShut {NoStop}%
\bibitem [{\citenamefont {{Deng}}\ \emph {et~al.}(2012)\citenamefont {{Deng}},
  \citenamefont {{Yu}}, \citenamefont {{Huang}}, \citenamefont {{Larsson}},
  \citenamefont {{Caroff}},\ and\ \citenamefont {{Xu}}}]{Deng2012}%
  \BibitemOpen
  \bibfield  {author} {\bibinfo {author} {\bibfnamefont {M.~T.}\ \bibnamefont
  {{Deng}}}, \bibinfo {author} {\bibfnamefont {C.~L.}\ \bibnamefont {{Yu}}},
  \bibinfo {author} {\bibfnamefont {G.~Y.}\ \bibnamefont {{Huang}}}, \bibinfo
  {author} {\bibfnamefont {M.}~\bibnamefont {{Larsson}}}, \bibinfo {author}
  {\bibfnamefont {P.}~\bibnamefont {{Caroff}}}, \ and\ \bibinfo {author}
  {\bibfnamefont {H.~Q.}\ \bibnamefont {{Xu}}},\ }\href {\doibase
  10.1021/nl303758w} {\bibfield  {journal} {\bibinfo  {journal} {Nano Lett.}\
  }\textbf {\bibinfo {volume} {12}},\ \bibinfo {pages} {6414} (\bibinfo {year}
  {2012})}\BibitemShut {NoStop}%
\bibitem [{\citenamefont {{Finck}}\ \emph {et~al.}(2013)\citenamefont
  {{Finck}}, \citenamefont {{Van Harlingen}}, \citenamefont {{Mohseni}},
  \citenamefont {{Jung}},\ and\ \citenamefont {{Li}}}]{Fink2012}%
  \BibitemOpen
  \bibfield  {author} {\bibinfo {author} {\bibfnamefont {A.~D.~K.}\
  \bibnamefont {{Finck}}}, \bibinfo {author} {\bibfnamefont {D.~J.}\
  \bibnamefont {{Van Harlingen}}}, \bibinfo {author} {\bibfnamefont {P.~K.}\
  \bibnamefont {{Mohseni}}}, \bibinfo {author} {\bibfnamefont {K.}~\bibnamefont
  {{Jung}}}, \ and\ \bibinfo {author} {\bibfnamefont {X.}~\bibnamefont
  {{Li}}},\ }\href {\doibase 10.1103/PhysRevLett.110.126406} {\bibfield
  {journal} {\bibinfo  {journal} {\prl}\ }\textbf {\bibinfo {volume} {110}},\
  \bibinfo {pages} {126406} (\bibinfo {year} {2013})}\BibitemShut {NoStop}%
\bibitem [{\citenamefont {Churchill}\ \emph {et~al.}(2013)\citenamefont
  {Churchill}, \citenamefont {Fatemi}, \citenamefont {Grove-Rasmussen},
  \citenamefont {Deng}, \citenamefont {Caroff}, \citenamefont {Xu},\ and\
  \citenamefont {Marcus}}]{Churchill2013}%
  \BibitemOpen
  \bibfield  {author} {\bibinfo {author} {\bibfnamefont {H.~O.~H.}\
  \bibnamefont {Churchill}}, \bibinfo {author} {\bibfnamefont {V.}~\bibnamefont
  {Fatemi}}, \bibinfo {author} {\bibfnamefont {K.}~\bibnamefont
  {Grove-Rasmussen}}, \bibinfo {author} {\bibfnamefont {M.~T.}\ \bibnamefont
  {Deng}}, \bibinfo {author} {\bibfnamefont {P.}~\bibnamefont {Caroff}},
  \bibinfo {author} {\bibfnamefont {H.~Q.}\ \bibnamefont {Xu}}, \ and\ \bibinfo
  {author} {\bibfnamefont {C.~M.}\ \bibnamefont {Marcus}},\ }\href {\doibase
  10.1103/PhysRevB.87.241401} {\bibfield  {journal} {\bibinfo  {journal}
  {\prb}\ }\textbf {\bibinfo {volume} {87}},\ \bibinfo {pages} {241401}
  (\bibinfo {year} {2013})}\BibitemShut {NoStop}%
\bibitem [{\citenamefont {Deng}\ \emph {et~al.}(2014)\citenamefont {Deng},
  \citenamefont {Yu}, \citenamefont {Huang}, \citenamefont {Larsson},
  \citenamefont {Caroff},\ and\ \citenamefont {Xu}}]{Deng2014}%
  \BibitemOpen
  \bibfield  {author} {\bibinfo {author} {\bibfnamefont {M.~T.}\ \bibnamefont
  {Deng}}, \bibinfo {author} {\bibfnamefont {C.~L.}\ \bibnamefont {Yu}},
  \bibinfo {author} {\bibfnamefont {G.~Y.}\ \bibnamefont {Huang}}, \bibinfo
  {author} {\bibfnamefont {M.}~\bibnamefont {Larsson}}, \bibinfo {author}
  {\bibfnamefont {P.}~\bibnamefont {Caroff}}, \ and\ \bibinfo {author}
  {\bibfnamefont {H.~Q.}\ \bibnamefont {Xu}},\ }\href {\doibase
  doi:10.1038/srep07261} {\bibfield  {journal} {\bibinfo  {journal} {Scientific
  Reports}\ }\textbf {\bibinfo {volume} {4}},\ \bibinfo {pages} {7261}
  (\bibinfo {year} {2014})}\BibitemShut {NoStop}%
\bibitem [{\citenamefont {{Higginbotham}}\ \emph {et~al.}(2015)\citenamefont
  {{Higginbotham}}, \citenamefont {{Albrecht}}, \citenamefont {{Kirsanskas}},
  \citenamefont {{Chang}}, \citenamefont {{Kuemmeth}}, \citenamefont
  {{Krogstrup}}, \citenamefont {{Jespersen}}, \citenamefont {{Nygard}},
  \citenamefont {{Flensberg}},\ and\ \citenamefont
  {{Marcus}}}]{Higginbotham15}%
  \BibitemOpen
  \bibfield  {author} {\bibinfo {author} {\bibfnamefont {A.~P.}\ \bibnamefont
  {{Higginbotham}}}, \bibinfo {author} {\bibfnamefont {S.~M.}\ \bibnamefont
  {{Albrecht}}}, \bibinfo {author} {\bibfnamefont {G.}~\bibnamefont
  {{Kirsanskas}}}, \bibinfo {author} {\bibfnamefont {W.}~\bibnamefont
  {{Chang}}}, \bibinfo {author} {\bibfnamefont {F.}~\bibnamefont {{Kuemmeth}}},
  \bibinfo {author} {\bibfnamefont {P.}~\bibnamefont {{Krogstrup}}}, \bibinfo
  {author} {\bibfnamefont {T.~S.}\ \bibnamefont {{Jespersen}}}, \bibinfo
  {author} {\bibfnamefont {J.}~\bibnamefont {{Nygard}}}, \bibinfo {author}
  {\bibfnamefont {K.}~\bibnamefont {{Flensberg}}}, \ and\ \bibinfo {author}
  {\bibfnamefont {C.~M.}\ \bibnamefont {{Marcus}}},\ }\href@noop {} {\bibfield
  {journal} {\bibinfo  {journal} {Nature Physics}\ }\textbf {\bibinfo {volume}
  {11}},\ \bibinfo {pages} {1017} (\bibinfo {year} {2015})},\ \Eprint
  {http://arxiv.org/abs/1501.05155} {arXiv:1501.05155 [cond-mat.mes-hall]}
  \BibitemShut {NoStop}%
\bibitem [{\citenamefont {Albrecht}\ \emph {et~al.}(2016)\citenamefont
  {Albrecht}, \citenamefont {Higginbotham}, \citenamefont {Madsen},
  \citenamefont {Kuemmeth}, \citenamefont {Jespersen}, \citenamefont
  {Nyg{\aa}rd}, \citenamefont {Krogstrup},\ and\ \citenamefont
  {Marcus}}]{albrecht2015}%
  \BibitemOpen
  \bibfield  {author} {\bibinfo {author} {\bibfnamefont {S.~M.}\ \bibnamefont
  {Albrecht}}, \bibinfo {author} {\bibfnamefont {A.~P.}\ \bibnamefont
  {Higginbotham}}, \bibinfo {author} {\bibfnamefont {M.}~\bibnamefont
  {Madsen}}, \bibinfo {author} {\bibfnamefont {F.}~\bibnamefont {Kuemmeth}},
  \bibinfo {author} {\bibfnamefont {T.~S.}\ \bibnamefont {Jespersen}}, \bibinfo
  {author} {\bibfnamefont {J.}~\bibnamefont {Nyg{\aa}rd}}, \bibinfo {author}
  {\bibfnamefont {P.}~\bibnamefont {Krogstrup}}, \ and\ \bibinfo {author}
  {\bibfnamefont {C.~M.}\ \bibnamefont {Marcus}},\ }\href {\doibase
  10.1038/nature17162} {\bibfield  {journal} {\bibinfo  {journal} {Nature}\
  }\textbf {\bibinfo {volume} {531}},\ \bibinfo {pages} {206} (\bibinfo {year}
  {2016})}\BibitemShut {NoStop}%
\bibitem [{\citenamefont {Zhang}\ \emph {et~al.}(2016)\citenamefont {Zhang},
  \citenamefont {Gul}, \citenamefont {Conesa-Boj}, \citenamefont {Zuo},
  \citenamefont {Mourik}, \citenamefont {de~Vries}, \citenamefont {van Veen},
  \citenamefont {van Woerkom}, \citenamefont {Nowak}, \citenamefont {Wimmer},
  \citenamefont {Car}, \citenamefont {Plissard}, \citenamefont {Bakkers},
  \citenamefont {Quintero$-$Perez}, \citenamefont {Goswami}, \citenamefont
  {Watanabe}, \citenamefont {Taniguchi},\ and\ \citenamefont
  {Kouwenhoven}}]{HaoZhang16}%
  \BibitemOpen
  \bibfield  {author} {\bibinfo {author} {\bibfnamefont {H.}~\bibnamefont
  {Zhang}}, \bibinfo {author} {\bibfnamefont {O.}~\bibnamefont {Gul}}, \bibinfo
  {author} {\bibfnamefont {S.}~\bibnamefont {Conesa-Boj}}, \bibinfo {author}
  {\bibfnamefont {K.}~\bibnamefont {Zuo}}, \bibinfo {author} {\bibfnamefont
  {V.}~\bibnamefont {Mourik}}, \bibinfo {author} {\bibfnamefont {F.~K.}\
  \bibnamefont {de~Vries}}, \bibinfo {author} {\bibfnamefont {J.}~\bibnamefont
  {van Veen}}, \bibinfo {author} {\bibfnamefont {D.~J.}\ \bibnamefont {van
  Woerkom}}, \bibinfo {author} {\bibfnamefont {M.~P.}\ \bibnamefont {Nowak}},
  \bibinfo {author} {\bibfnamefont {M.}~\bibnamefont {Wimmer}}, \bibinfo
  {author} {\bibfnamefont {D.}~\bibnamefont {Car}}, \bibinfo {author}
  {\bibfnamefont {S.}~\bibnamefont {Plissard}}, \bibinfo {author}
  {\bibfnamefont {E.~P. A.~M.}\ \bibnamefont {Bakkers}}, \bibinfo {author}
  {\bibfnamefont {M.}~\bibnamefont {Quintero$-$Perez}}, \bibinfo {author}
  {\bibfnamefont {S.}~\bibnamefont {Goswami}}, \bibinfo {author} {\bibfnamefont
  {K.}~\bibnamefont {Watanabe}}, \bibinfo {author} {\bibfnamefont
  {T.}~\bibnamefont {Taniguchi}}, \ and\ \bibinfo {author} {\bibfnamefont
  {L.~P.}\ \bibnamefont {Kouwenhoven}},\ }\href@noop {} {\bibfield  {journal}
  {\bibinfo  {journal} {arXiv:1603.04069}\ } (\bibinfo {year}
  {2016})}\BibitemShut {NoStop}%
\bibitem [{\citenamefont {Deng}\ \emph {et~al.}(2016)\citenamefont {Deng},
  \citenamefont {Vaitiekenas}, \citenamefont {Hansen}, \citenamefont {Danon},
  \citenamefont {Leijnse}, \citenamefont {Flensberg}, \citenamefont {Nyg{\r
  a}rd}, \citenamefont {Krogstrup},\ and\ \citenamefont {Marcus}}]{Deng16}%
  \BibitemOpen
  \bibfield  {author} {\bibinfo {author} {\bibfnamefont {M.~T.}\ \bibnamefont
  {Deng}}, \bibinfo {author} {\bibfnamefont {S.}~\bibnamefont {Vaitiekenas}},
  \bibinfo {author} {\bibfnamefont {E.~B.}\ \bibnamefont {Hansen}}, \bibinfo
  {author} {\bibfnamefont {J.}~\bibnamefont {Danon}}, \bibinfo {author}
  {\bibfnamefont {M.}~\bibnamefont {Leijnse}}, \bibinfo {author} {\bibfnamefont
  {K.}~\bibnamefont {Flensberg}}, \bibinfo {author} {\bibfnamefont
  {J.}~\bibnamefont {Nyg{\r a}rd}}, \bibinfo {author} {\bibfnamefont
  {P.}~\bibnamefont {Krogstrup}}, \ and\ \bibinfo {author} {\bibfnamefont
  {C.~M.}\ \bibnamefont {Marcus}},\ }\href {\doibase 10.1126/science.aaf3961}
  {\bibfield  {journal} {\bibinfo  {journal} {Science}\ }\textbf {\bibinfo
  {volume} {354}},\ \bibinfo {pages} {1557} (\bibinfo {year}
  {2016})}\BibitemShut {NoStop}%
\bibitem [{\citenamefont {{Hyart}}\ \emph {et~al.}(2013)\citenamefont
  {{Hyart}}, \citenamefont {{van Heck}}, \citenamefont {{Fulga}}, \citenamefont
  {{Burrello}}, \citenamefont {{Akhmerov}},\ and\ \citenamefont
  {{Beenakker}}}]{Hyart13}%
  \BibitemOpen
  \bibfield  {author} {\bibinfo {author} {\bibfnamefont {T.}~\bibnamefont
  {{Hyart}}}, \bibinfo {author} {\bibfnamefont {B.}~\bibnamefont {{van Heck}}},
  \bibinfo {author} {\bibfnamefont {I.~C.}\ \bibnamefont {{Fulga}}}, \bibinfo
  {author} {\bibfnamefont {M.}~\bibnamefont {{Burrello}}}, \bibinfo {author}
  {\bibfnamefont {A.~R.}\ \bibnamefont {{Akhmerov}}}, \ and\ \bibinfo {author}
  {\bibfnamefont {C.~W.~J.}\ \bibnamefont {{Beenakker}}},\ }\href {\doibase
  10.1103/PhysRevB.88.035121} {\bibfield  {journal} {\bibinfo  {journal}
  {\prb}\ }\textbf {\bibinfo {volume} {88}},\ \bibinfo {eid} {035121} (\bibinfo
  {year} {2013})},\ \Eprint {http://arxiv.org/abs/1303.4379} {arXiv:1303.4379
  [quant-ph]} \BibitemShut {NoStop}%
\bibitem [{\citenamefont {{Clarke}}\ \emph {et~al.}(2016)\citenamefont
  {{Clarke}}, \citenamefont {{Sau}},\ and\ \citenamefont {{Das
  Sarma}}}]{Clarke16}%
  \BibitemOpen
  \bibfield  {author} {\bibinfo {author} {\bibfnamefont {D.~J.}\ \bibnamefont
  {{Clarke}}}, \bibinfo {author} {\bibfnamefont {J.~D.}\ \bibnamefont {{Sau}}},
  \ and\ \bibinfo {author} {\bibfnamefont {S.}~\bibnamefont {{Das Sarma}}},\
  }\href {\doibase 10.1103/PhysRevX.6.021005} {\bibfield  {journal} {\bibinfo
  {journal} {Physical Review X}\ }\textbf {\bibinfo {volume} {6}},\ \bibinfo
  {eid} {021005} (\bibinfo {year} {2016})},\ \Eprint
  {http://arxiv.org/abs/1510.00007} {arXiv:1510.00007 [quant-ph]} \BibitemShut
  {NoStop}%
\bibitem [{\citenamefont {{Aasen}}\ \emph {et~al.}(2016)\citenamefont
  {{Aasen}}, \citenamefont {{Lee}}, \citenamefont {{Karzig}},\ and\
  \citenamefont {{Alicea}}}]{Aasen16}%
  \BibitemOpen
  \bibfield  {author} {\bibinfo {author} {\bibfnamefont {D.}~\bibnamefont
  {{Aasen}}}, \bibinfo {author} {\bibfnamefont {S.-P.}\ \bibnamefont {{Lee}}},
  \bibinfo {author} {\bibfnamefont {T.}~\bibnamefont {{Karzig}}}, \ and\
  \bibinfo {author} {\bibfnamefont {J.}~\bibnamefont {{Alicea}}},\ }\href
  {\doibase 10.1103/PhysRevB.94.165113} {\bibfield  {journal} {\bibinfo
  {journal} {\prb}\ }\textbf {\bibinfo {volume} {94}},\ \bibinfo {eid} {165113}
  (\bibinfo {year} {2016})},\ \Eprint {http://arxiv.org/abs/1606.09255}
  {arXiv:1606.09255 [cond-mat.mes-hall]} \BibitemShut {NoStop}%
\bibitem [{\citenamefont {{Landau}}\ \emph {et~al.}(2016)\citenamefont
  {{Landau}}, \citenamefont {{Plugge}}, \citenamefont {{Sela}}, \citenamefont
  {{Altland}}, \citenamefont {{Albrecht}},\ and\ \citenamefont
  {{Egger}}}]{Landau16}%
  \BibitemOpen
  \bibfield  {author} {\bibinfo {author} {\bibfnamefont {L.~A.}\ \bibnamefont
  {{Landau}}}, \bibinfo {author} {\bibfnamefont {S.}~\bibnamefont {{Plugge}}},
  \bibinfo {author} {\bibfnamefont {E.}~\bibnamefont {{Sela}}}, \bibinfo
  {author} {\bibfnamefont {A.}~\bibnamefont {{Altland}}}, \bibinfo {author}
  {\bibfnamefont {S.~M.}\ \bibnamefont {{Albrecht}}}, \ and\ \bibinfo {author}
  {\bibfnamefont {R.}~\bibnamefont {{Egger}}},\ }\href {\doibase
  10.1103/PhysRevLett.116.050501} {\bibfield  {journal} {\bibinfo  {journal}
  {Physical Review Letters}\ }\textbf {\bibinfo {volume} {116}},\ \bibinfo
  {eid} {050501} (\bibinfo {year} {2016})},\ \Eprint
  {http://arxiv.org/abs/1509.05345} {arXiv:1509.05345} \BibitemShut {NoStop}%
\bibitem [{\citenamefont {{Plugge}}\ \emph
  {et~al.}(2016{\natexlab{a}})\citenamefont {{Plugge}}, \citenamefont
  {{Landau}}, \citenamefont {{Sela}}, \citenamefont {{Altland}}, \citenamefont
  {{Flensberg}},\ and\ \citenamefont {{Egger}}}]{Plugge16a}%
  \BibitemOpen
  \bibfield  {author} {\bibinfo {author} {\bibfnamefont {S.}~\bibnamefont
  {{Plugge}}}, \bibinfo {author} {\bibfnamefont {L.~A.}\ \bibnamefont
  {{Landau}}}, \bibinfo {author} {\bibfnamefont {E.}~\bibnamefont {{Sela}}},
  \bibinfo {author} {\bibfnamefont {A.}~\bibnamefont {{Altland}}}, \bibinfo
  {author} {\bibfnamefont {K.}~\bibnamefont {{Flensberg}}}, \ and\ \bibinfo
  {author} {\bibfnamefont {R.}~\bibnamefont {{Egger}}},\ }\href@noop {}
  {\enquote {\bibinfo {title} {{Roadmap to Majorana surface codes}},}\ }
  (\bibinfo {year} {2016}{\natexlab{a}}),\ \Eprint
  {http://arxiv.org/abs/arXiv:1606.08408} {arXiv:1606.08408} \BibitemShut
  {NoStop}%
\bibitem [{\citenamefont {{Plugge}}\ \emph
  {et~al.}(2016{\natexlab{b}})\citenamefont {{Plugge}}, \citenamefont
  {{Rasmussen}}, \citenamefont {{Egger}},\ and\ \citenamefont
  {{Flensberg}}}]{Plugge16}%
  \BibitemOpen
  \bibfield  {author} {\bibinfo {author} {\bibfnamefont {S.}~\bibnamefont
  {{Plugge}}}, \bibinfo {author} {\bibfnamefont {A.}~\bibnamefont
  {{Rasmussen}}}, \bibinfo {author} {\bibfnamefont {R.}~\bibnamefont
  {{Egger}}}, \ and\ \bibinfo {author} {\bibfnamefont {K.}~\bibnamefont
  {{Flensberg}}},\ }\href@noop {} {\bibfield  {journal} {\bibinfo  {journal}
  {ArXiv e-prints}\ } (\bibinfo {year} {2016}{\natexlab{b}})},\ \Eprint
  {http://arxiv.org/abs/1609.01697} {arXiv:1609.01697 [cond-mat.mes-hall]}
  \BibitemShut {NoStop}%
\bibitem [{\citenamefont {{Karzig}}\ \emph {et~al.}(2016)\citenamefont
  {{Karzig}}, \citenamefont {{Knapp}}, \citenamefont {{Lutchyn}}, \citenamefont
  {{Bonderson}}, \citenamefont {{Hastings}}, \citenamefont {{Nayak}},
  \citenamefont {{Alicea}}, \citenamefont {{Flensberg}}, \citenamefont
  {{Plugge}}, \citenamefont {{Oreg}}, \citenamefont {{Marcus}},\ and\
  \citenamefont {{Freedman}}}]{Karzig16}%
  \BibitemOpen
  \bibfield  {author} {\bibinfo {author} {\bibfnamefont {T.}~\bibnamefont
  {{Karzig}}}, \bibinfo {author} {\bibfnamefont {C.}~\bibnamefont {{Knapp}}},
  \bibinfo {author} {\bibfnamefont {R.}~\bibnamefont {{Lutchyn}}}, \bibinfo
  {author} {\bibfnamefont {P.}~\bibnamefont {{Bonderson}}}, \bibinfo {author}
  {\bibfnamefont {M.}~\bibnamefont {{Hastings}}}, \bibinfo {author}
  {\bibfnamefont {C.}~\bibnamefont {{Nayak}}}, \bibinfo {author} {\bibfnamefont
  {J.}~\bibnamefont {{Alicea}}}, \bibinfo {author} {\bibfnamefont
  {K.}~\bibnamefont {{Flensberg}}}, \bibinfo {author} {\bibfnamefont
  {S.}~\bibnamefont {{Plugge}}}, \bibinfo {author} {\bibfnamefont
  {Y.}~\bibnamefont {{Oreg}}}, \bibinfo {author} {\bibfnamefont
  {C.}~\bibnamefont {{Marcus}}}, \ and\ \bibinfo {author} {\bibfnamefont
  {M.~H.}\ \bibnamefont {{Freedman}}},\ }\href@noop {} {\bibfield  {journal}
  {\bibinfo  {journal} {ArXiv e-prints}\ } (\bibinfo {year} {2016})},\ \Eprint
  {http://arxiv.org/abs/1610.05289} {arXiv:1610.05289 [cond-mat.mes-hall]}
  \BibitemShut {NoStop}%
\bibitem [{\citenamefont {Matveev}\ \emph {et~al.}(1994)\citenamefont
  {Matveev}, \citenamefont {Glazman},\ and\ \citenamefont
  {Shekhter}}]{Glazman'94}%
  \BibitemOpen
  \bibfield  {author} {\bibinfo {author} {\bibfnamefont {K.~A.}\ \bibnamefont
  {Matveev}}, \bibinfo {author} {\bibfnamefont {L.~I.}\ \bibnamefont
  {Glazman}}, \ and\ \bibinfo {author} {\bibfnamefont {R.~I.}\ \bibnamefont
  {Shekhter}},\ }\href {\doibase 10.1142/S0217984994001011} {\bibfield
  {journal} {\bibinfo  {journal} {Modern Physics Letters B}\ }\textbf {\bibinfo
  {volume} {08}},\ \bibinfo {pages} {1007} (\bibinfo {year}
  {1994})}\BibitemShut {NoStop}%
\bibitem [{\citenamefont {von Delft}\ and\ \citenamefont
  {Ralph}(2001)}]{vonDelft'01}%
  \BibitemOpen
  \bibfield  {author} {\bibinfo {author} {\bibfnamefont {J.}~\bibnamefont {von
  Delft}}\ and\ \bibinfo {author} {\bibfnamefont {D.}~\bibnamefont {Ralph}},\
  }\href {\doibase http://dx.doi.org/10.1016/S0370-1573(00)00099-5} {\bibfield
  {journal} {\bibinfo  {journal} {Physics Reports}\ }\textbf {\bibinfo {volume}
  {345}},\ \bibinfo {pages} {61 } (\bibinfo {year} {2001})}\BibitemShut
  {NoStop}%
\bibitem [{\citenamefont {Fu}(2010)}]{Fu2010}%
  \BibitemOpen
  \bibfield  {author} {\bibinfo {author} {\bibfnamefont {L.}~\bibnamefont
  {Fu}},\ }\href {\doibase 10.1103/PhysRevLett.104.056402} {\bibfield
  {journal} {\bibinfo  {journal} {Phys. Rev. Lett.}\ }\textbf {\bibinfo
  {volume} {104}},\ \bibinfo {pages} {056402} (\bibinfo {year}
  {2010})}\BibitemShut {NoStop}%
\bibitem [{\citenamefont {{van Heck}}\ \emph {et~al.}()\citenamefont {{van
  Heck}}, \citenamefont {{Lutchyn}},\ and\ \citenamefont
  {{Glazman}}}]{vanHeck16}%
  \BibitemOpen
  \bibfield  {author} {\bibinfo {author} {\bibfnamefont {B.}~\bibnamefont {{van
  Heck}}}, \bibinfo {author} {\bibfnamefont {R.~M.}\ \bibnamefont {{Lutchyn}}},
  \ and\ \bibinfo {author} {\bibfnamefont {L.~I.}\ \bibnamefont {{Glazman}}},\
  }\href@noop {} {\bibfield  {journal} {\bibinfo  {journal} {ArXiv e-prints}\
  }}\Eprint {http://arxiv.org/abs/1603.08258} {arXiv:1603.08258} \BibitemShut
  {NoStop}%
\bibitem [{\citenamefont {{Lutchyn}}\ and\ \citenamefont
  {{Glazman}}(2017)}]{Lutchyn2017}%
  \BibitemOpen
  \bibfield  {author} {\bibinfo {author} {\bibfnamefont {R.~M.}\ \bibnamefont
  {{Lutchyn}}}\ and\ \bibinfo {author} {\bibfnamefont {L.~I.}\ \bibnamefont
  {{Glazman}}},\ }\href@noop {} {\bibfield  {journal} {\bibinfo  {journal}
  {ArXiv e-prints}\ } (\bibinfo {year} {2017})},\ \Eprint
  {http://arxiv.org/abs/1701.00184} {arXiv:1701.00184 [cond-mat.supr-con]}
  \BibitemShut {NoStop}%
\bibitem [{\citenamefont {{Fidkowski}}\ and\ \citenamefont
  {{Kitaev}}(2011)}]{Fidkowski11a}%
  \BibitemOpen
  \bibfield  {author} {\bibinfo {author} {\bibfnamefont {L.}~\bibnamefont
  {{Fidkowski}}}\ and\ \bibinfo {author} {\bibfnamefont {A.}~\bibnamefont
  {{Kitaev}}},\ }\href {\doibase 10.1103/PhysRevB.83.075103} {\bibfield
  {journal} {\bibinfo  {journal} {\prb}\ }\textbf {\bibinfo {volume} {83}},\
  \bibinfo {pages} {075103} (\bibinfo {year} {2011})},\ \Eprint
  {http://arxiv.org/abs/1008.4138} {arXiv:1008.4138 [cond-mat.str-el]}
  \BibitemShut {NoStop}%
\bibitem [{\citenamefont {{Clarke}}\ \emph {et~al.}(2013)\citenamefont
  {{Clarke}}, \citenamefont {{Alicea}},\ and\ \citenamefont
  {{Shtengel}}}]{Clarke2013a}%
  \BibitemOpen
  \bibfield  {author} {\bibinfo {author} {\bibfnamefont {D.~J.}\ \bibnamefont
  {{Clarke}}}, \bibinfo {author} {\bibfnamefont {J.}~\bibnamefont {{Alicea}}},
  \ and\ \bibinfo {author} {\bibfnamefont {K.}~\bibnamefont {{Shtengel}}},\
  }\href {\doibase 10.1038/ncomms2340} {\bibfield  {journal} {\bibinfo
  {journal} {Nature Communications}\ }\textbf {\bibinfo {volume} {4}},\
  \bibinfo {eid} {1348} (\bibinfo {year} {2013})},\ \Eprint
  {http://arxiv.org/abs/1204.5479} {arXiv:1204.5479 [cond-mat.str-el]}
  \BibitemShut {NoStop}%
\bibitem [{\citenamefont {{Lindner}}\ \emph {et~al.}(2012)\citenamefont
  {{Lindner}}, \citenamefont {{Berg}}, \citenamefont {{Refael}},\ and\
  \citenamefont {{Stern}}}]{Lindner2012}%
  \BibitemOpen
  \bibfield  {author} {\bibinfo {author} {\bibfnamefont {N.~H.}\ \bibnamefont
  {{Lindner}}}, \bibinfo {author} {\bibfnamefont {E.}~\bibnamefont {{Berg}}},
  \bibinfo {author} {\bibfnamefont {G.}~\bibnamefont {{Refael}}}, \ and\
  \bibinfo {author} {\bibfnamefont {A.}~\bibnamefont {{Stern}}},\ }\href
  {\doibase 10.1103/PhysRevX.2.041002} {\bibfield  {journal} {\bibinfo
  {journal} {Physical Review X}\ }\textbf {\bibinfo {volume} {2}},\ \bibinfo
  {eid} {041002} (\bibinfo {year} {2012})},\ \Eprint
  {http://arxiv.org/abs/1204.5733} {arXiv:1204.5733 [cond-mat.mes-hall]}
  \BibitemShut {NoStop}%
\bibitem [{\citenamefont {{Cheng}}(2012)}]{Cheng2012}%
  \BibitemOpen
  \bibfield  {author} {\bibinfo {author} {\bibfnamefont {M.}~\bibnamefont
  {{Cheng}}},\ }\href {\doibase 10.1103/PhysRevB.86.195126} {\bibfield
  {journal} {\bibinfo  {journal} {\prb}\ }\textbf {\bibinfo {volume} {86}},\
  \bibinfo {eid} {195126} (\bibinfo {year} {2012})},\ \Eprint
  {http://arxiv.org/abs/1204.6084} {arXiv:1204.6084 [cond-mat.str-el]}
  \BibitemShut {NoStop}%
\bibitem [{\citenamefont {Barkeshli}(2016)}]{Barkeshli16}%
  \BibitemOpen
  \bibfield  {author} {\bibinfo {author} {\bibfnamefont {M.}~\bibnamefont
  {Barkeshli}},\ }\href {\doibase 10.1103/PhysRevLett.117.096803} {\bibfield
  {journal} {\bibinfo  {journal} {Phys. Rev. Lett.}\ }\textbf {\bibinfo
  {volume} {117}},\ \bibinfo {pages} {096803} (\bibinfo {year}
  {2016})}\BibitemShut {NoStop}%
\bibitem [{\citenamefont {Clarke}\ \emph {et~al.}(2014)\citenamefont {Clarke},
  \citenamefont {Alicea},\ and\ \citenamefont {Shtengel}}]{Clarke13}%
  \BibitemOpen
  \bibfield  {author} {\bibinfo {author} {\bibfnamefont {D.~J.}\ \bibnamefont
  {Clarke}}, \bibinfo {author} {\bibfnamefont {J.}~\bibnamefont {Alicea}}, \
  and\ \bibinfo {author} {\bibfnamefont {K.}~\bibnamefont {Shtengel}},\ }\href
  {\doibase 10.1038/nphys3114} {\bibfield  {journal} {\bibinfo  {journal}
  {Nature Physics}\ }\textbf {\bibinfo {volume} {10}},\ \bibinfo {pages} {877}
  (\bibinfo {year} {2014})},\ \Eprint {http://arxiv.org/abs/1312.6123}
  {arXiv:1312.6123 [cond-mat.str-el]} \BibitemShut {NoStop}%
\bibitem [{\citenamefont {Zazunov}\ \emph {et~al.}(2011)\citenamefont
  {Zazunov}, \citenamefont {Yeyati},\ and\ \citenamefont {Egger}}]{Zazunov11}%
  \BibitemOpen
  \bibfield  {author} {\bibinfo {author} {\bibfnamefont {A.}~\bibnamefont
  {Zazunov}}, \bibinfo {author} {\bibfnamefont {A.~L.}\ \bibnamefont {Yeyati}},
  \ and\ \bibinfo {author} {\bibfnamefont {R.}~\bibnamefont {Egger}},\ }\href
  {\doibase 10.1103/PhysRevB.84.165440} {\bibfield  {journal} {\bibinfo
  {journal} {Phys. Rev. B}\ }\textbf {\bibinfo {volume} {84}},\ \bibinfo
  {pages} {165440} (\bibinfo {year} {2011})},\ \Eprint
  {http://arxiv.org/abs/arXiv:1108.4308} {arXiv:1108.4308} \BibitemShut
  {NoStop}%
\bibitem [{\citenamefont {{Lutchyn}}\ \emph {et~al.}(2016)\citenamefont
  {{Lutchyn}}, \citenamefont {{Flensberg}},\ and\ \citenamefont
  {{Glazman}}}]{Lutchyn16}%
  \BibitemOpen
  \bibfield  {author} {\bibinfo {author} {\bibfnamefont {R.~M.}\ \bibnamefont
  {{Lutchyn}}}, \bibinfo {author} {\bibfnamefont {K.}~\bibnamefont
  {{Flensberg}}}, \ and\ \bibinfo {author} {\bibfnamefont {L.~I.}\ \bibnamefont
  {{Glazman}}},\ }\href {\doibase 10.1103/PhysRevB.94.125407} {\bibfield
  {journal} {\bibinfo  {journal} {\prb}\ }\textbf {\bibinfo {volume} {94}},\
  \bibinfo {eid} {125407} (\bibinfo {year} {2016})},\ \Eprint
  {http://arxiv.org/abs/1606.06756} {arXiv:1606.06756 [cond-mat.mes-hall]}
  \BibitemShut {NoStop}%
\bibitem [{\citenamefont {{Iftikhar}}\ \emph {et~al.}(2015)\citenamefont
  {{Iftikhar}}, \citenamefont {{Jezouin}}, \citenamefont {{Anthore}},
  \citenamefont {{Gennser}}, \citenamefont {{Parmentier}}, \citenamefont
  {{Cavanna}},\ and\ \citenamefont {{Pierre}}}]{Iftikhar'15}%
  \BibitemOpen
  \bibfield  {author} {\bibinfo {author} {\bibfnamefont {Z.}~\bibnamefont
  {{Iftikhar}}}, \bibinfo {author} {\bibfnamefont {S.}~\bibnamefont
  {{Jezouin}}}, \bibinfo {author} {\bibfnamefont {A.}~\bibnamefont
  {{Anthore}}}, \bibinfo {author} {\bibfnamefont {U.}~\bibnamefont
  {{Gennser}}}, \bibinfo {author} {\bibfnamefont {F.~D.}\ \bibnamefont
  {{Parmentier}}}, \bibinfo {author} {\bibfnamefont {A.}~\bibnamefont
  {{Cavanna}}}, \ and\ \bibinfo {author} {\bibfnamefont {F.}~\bibnamefont
  {{Pierre}}},\ }\href {\doibase 10.1038/nature15384} {\bibfield  {journal}
  {\bibinfo  {journal} {\nat}\ }\textbf {\bibinfo {volume} {526}},\ \bibinfo
  {pages} {233} (\bibinfo {year} {2015})},\ \Eprint
  {http://arxiv.org/abs/1602.02056} {arXiv:1602.02056 [cond-mat.mes-hall]}
  \BibitemShut {NoStop}%
\bibitem [{\citenamefont {{Jezouin}}\ \emph {et~al.}(2016)\citenamefont
  {{Jezouin}}, \citenamefont {{Iftikhar}}, \citenamefont {{Anthore}},
  \citenamefont {{Parmentier}}, \citenamefont {{Gennser}}, \citenamefont
  {{Cavanna}}, \citenamefont {{Ouerghi}}, \citenamefont {{Levkivskyi}},
  \citenamefont {{Idrisov}}, \citenamefont {{Sukhorukov}}, \citenamefont
  {{Glazman}},\ and\ \citenamefont {{Pierre}}}]{Jezouin'16}%
  \BibitemOpen
  \bibfield  {author} {\bibinfo {author} {\bibfnamefont {S.}~\bibnamefont
  {{Jezouin}}}, \bibinfo {author} {\bibfnamefont {Z.}~\bibnamefont
  {{Iftikhar}}}, \bibinfo {author} {\bibfnamefont {A.}~\bibnamefont
  {{Anthore}}}, \bibinfo {author} {\bibfnamefont {F.~D.}\ \bibnamefont
  {{Parmentier}}}, \bibinfo {author} {\bibfnamefont {U.}~\bibnamefont
  {{Gennser}}}, \bibinfo {author} {\bibfnamefont {A.}~\bibnamefont
  {{Cavanna}}}, \bibinfo {author} {\bibfnamefont {A.}~\bibnamefont
  {{Ouerghi}}}, \bibinfo {author} {\bibfnamefont {I.~P.}\ \bibnamefont
  {{Levkivskyi}}}, \bibinfo {author} {\bibfnamefont {E.}~\bibnamefont
  {{Idrisov}}}, \bibinfo {author} {\bibfnamefont {E.~V.}\ \bibnamefont
  {{Sukhorukov}}}, \bibinfo {author} {\bibfnamefont {L.~I.}\ \bibnamefont
  {{Glazman}}}, \ and\ \bibinfo {author} {\bibfnamefont {F.}~\bibnamefont
  {{Pierre}}},\ }\href {\doibase 10.1038/nature19072} {\bibfield  {journal}
  {\bibinfo  {journal} {\nat}\ }\textbf {\bibinfo {volume} {536}},\ \bibinfo
  {pages} {58} (\bibinfo {year} {2016})},\ \Eprint
  {http://arxiv.org/abs/1609.08910} {arXiv:1609.08910 [cond-mat.mes-hall]}
  \BibitemShut {NoStop}%
\bibitem [{\citenamefont {Wen}\ and\ \citenamefont {Zee}(1992)}]{Wen1992a}%
  \BibitemOpen
  \bibfield  {author} {\bibinfo {author} {\bibfnamefont {X.~G.}\ \bibnamefont
  {Wen}}\ and\ \bibinfo {author} {\bibfnamefont {A.}~\bibnamefont {Zee}},\
  }\href {\doibase 10.1103/PhysRevB.46.2290} {\bibfield  {journal} {\bibinfo
  {journal} {Phys. Rev. B}\ }\textbf {\bibinfo {volume} {46}},\ \bibinfo
  {pages} {2290} (\bibinfo {year} {1992})}\BibitemShut {NoStop}%
\bibitem [{Note1()}]{Note1}%
  \BibitemOpen
  \bibinfo {note} {Here we have chosen a gauge in which all the currents in the
  system are flowing along the boundary of the quantum Hall regions that lies
  between the superconductor and the four leads, while none flows between the
  leads along the uppermost or lowermost edge in Fig.~\ref
  {fig:device}}\BibitemShut {NoStop}%
\bibitem [{\citenamefont {Aleiner}\ \emph {et~al.}(2002)\citenamefont
  {Aleiner}, \citenamefont {Brouwer},\ and\ \citenamefont
  {Glazman}}]{Aleiner'02}%
  \BibitemOpen
  \bibfield  {author} {\bibinfo {author} {\bibfnamefont {I.}~\bibnamefont
  {Aleiner}}, \bibinfo {author} {\bibfnamefont {P.}~\bibnamefont {Brouwer}}, \
  and\ \bibinfo {author} {\bibfnamefont {L.}~\bibnamefont {Glazman}},\ }\href
  {\doibase http://dx.doi.org/10.1016/S0370-1573(01)00063-1} {\bibfield
  {journal} {\bibinfo  {journal} {Physics Reports}\ }\textbf {\bibinfo {volume}
  {358}},\ \bibinfo {pages} {309 } (\bibinfo {year} {2002})}\BibitemShut
  {NoStop}%
\bibitem [{Note2()}]{Note2}%
  \BibitemOpen
  \bibinfo {note} {In Appendix B, we present the cases $E_C\gg \Delta \gg D_c$
  and $\Delta \gg E_C$. Using similar RG analysis, one may show that in both
  these cases $G_1(T)=\nu e^2/h$ with a temperature-dependent correction
  scaling as $T^{3K_\rho -2}$ for $T\ll {\protect \rm min} \protect \{E_C,
  \Delta \protect \}$.}\BibitemShut {Stop}%
\bibitem [{Note3()}]{Note3}%
  \BibitemOpen
  \bibinfo {note} {Right- and left-moving representations are not independent,
  and one can equivalently write the Hamiltonian in terms of left-movers\cite
  {Clarke2013a,Mong14a,Mong14b}.}\BibitemShut {Stop}%
\bibitem [{\citenamefont {{Lutchyn}}\ and\ \citenamefont
  {{Skrabacz}}(2013)}]{Lutchyn2013}%
  \BibitemOpen
  \bibfield  {author} {\bibinfo {author} {\bibfnamefont {R.~M.}\ \bibnamefont
  {{Lutchyn}}}\ and\ \bibinfo {author} {\bibfnamefont {J.~H.}\ \bibnamefont
  {{Skrabacz}}},\ }\href {\doibase 10.1103/PhysRevB.88.024511} {\bibfield
  {journal} {\bibinfo  {journal} {\prb}\ }\textbf {\bibinfo {volume} {88}},\
  \bibinfo {eid} {024511} (\bibinfo {year} {2013})},\ \Eprint
  {http://arxiv.org/abs/1302.0289} {arXiv:1302.0289 [cond-mat.supr-con]}
  \BibitemShut {NoStop}%
\bibitem [{\citenamefont {{Mong}}\ \emph
  {et~al.}(2014{\natexlab{a}})\citenamefont {{Mong}}, \citenamefont {{Clarke}},
  \citenamefont {{Alicea}}, \citenamefont {{Lindner}}, \citenamefont
  {{Fendley}}, \citenamefont {{Nayak}}, \citenamefont {{Oreg}}, \citenamefont
  {{Stern}}, \citenamefont {{Berg}}, \citenamefont {{Shtengel}},\ and\
  \citenamefont {{Fisher}}}]{Mong14a}%
  \BibitemOpen
  \bibfield  {author} {\bibinfo {author} {\bibfnamefont {R.~S.~K.}\
  \bibnamefont {{Mong}}}, \bibinfo {author} {\bibfnamefont {D.~J.}\
  \bibnamefont {{Clarke}}}, \bibinfo {author} {\bibfnamefont {J.}~\bibnamefont
  {{Alicea}}}, \bibinfo {author} {\bibfnamefont {N.~H.}\ \bibnamefont
  {{Lindner}}}, \bibinfo {author} {\bibfnamefont {P.}~\bibnamefont
  {{Fendley}}}, \bibinfo {author} {\bibfnamefont {C.}~\bibnamefont {{Nayak}}},
  \bibinfo {author} {\bibfnamefont {Y.}~\bibnamefont {{Oreg}}}, \bibinfo
  {author} {\bibfnamefont {A.}~\bibnamefont {{Stern}}}, \bibinfo {author}
  {\bibfnamefont {E.}~\bibnamefont {{Berg}}}, \bibinfo {author} {\bibfnamefont
  {K.}~\bibnamefont {{Shtengel}}}, \ and\ \bibinfo {author} {\bibfnamefont
  {M.~P.~A.}\ \bibnamefont {{Fisher}}},\ }\href {\doibase
  10.1103/PhysRevX.4.011036} {\bibfield  {journal} {\bibinfo  {journal}
  {Physical Review X}\ }\textbf {\bibinfo {volume} {4}},\ \bibinfo {eid}
  {011036} (\bibinfo {year} {2014}{\natexlab{a}})},\ \Eprint
  {http://arxiv.org/abs/1307.4403} {arXiv:1307.4403 [cond-mat.str-el]}
  \BibitemShut {NoStop}%
\bibitem [{\citenamefont {{Mong}}\ \emph
  {et~al.}(2014{\natexlab{b}})\citenamefont {{Mong}}, \citenamefont {{Clarke}},
  \citenamefont {{Alicea}}, \citenamefont {{Lindner}},\ and\ \citenamefont
  {{Fendley}}}]{Mong14b}%
  \BibitemOpen
  \bibfield  {author} {\bibinfo {author} {\bibfnamefont {R.~S.~K.}\
  \bibnamefont {{Mong}}}, \bibinfo {author} {\bibfnamefont {D.~J.}\
  \bibnamefont {{Clarke}}}, \bibinfo {author} {\bibfnamefont {J.}~\bibnamefont
  {{Alicea}}}, \bibinfo {author} {\bibfnamefont {N.~H.}\ \bibnamefont
  {{Lindner}}}, \ and\ \bibinfo {author} {\bibfnamefont {P.}~\bibnamefont
  {{Fendley}}},\ }\href {\doibase 10.1088/1751-8113/47/45/452001} {\bibfield
  {journal} {\bibinfo  {journal} {Journal of Physics A Mathematical General}\
  }\textbf {\bibinfo {volume} {47}},\ \bibinfo {eid} {452001} (\bibinfo {year}
  {2014}{\natexlab{b}})},\ \Eprint {http://arxiv.org/abs/1406.0846}
  {arXiv:1406.0846 [cond-mat.stat-mech]} \BibitemShut {NoStop}%
\end{thebibliography}%
\end{document}